\documentclass[12pt]{article}

\begin{document}

\large \centerline{\bf Electric-Magnetic Duality in Massless  QED?
 }
\par
\vskip 0.5 truecm \normalsize
\begin{center}
{\bf C.~Ford}
\\
\vskip 0.5 truecm \it{
School of Mathematical Sciences\\
University College Dublin\\
 Belfield, Dublin 4\\Ireland\\
{\small\sf Christopher.Ford@ucd.ie}\\
}

\vskip 0.5 true cm

\end{center}

\vskip .7 truecm\normalsize
\begin{abstract}

The possibility that QED and recently developed non-Hermitian, or magnetic,
versions of QED are equivalent is considered.
Under this duality
 the Hamiltonians and anomalous axial currents
of the two theories are identified.  
A consequence of such a duality is that  particles
described by QED carry magnetic as well as  electric charges.
The proposal requires a vanishing zero bare 
fermion mass 
in both theories; Dirac mass terms are incompatible with the
 conservation of magnetic charge much as Majorana masses spoil the 
conservation of electric charge.
The physical spectrum comprises photons and
massless spin-${1\over2}$ particles carrying equal or opposite
electric and magnetic charges. The four particle states described by the Dirac 
fermion correspond to the four possible charge assignments of elementary dyons.
This scale invariant spectrum
indicates that the quantum field theory is finite.
The Johnson Baker Willey eigenvalue
equation   for the fine
structure constant in finite spinor QED
is interpreted as a Dirac-like charge quantisation condition for dyons.

\end{abstract}

\baselineskip=20pt
\bigskip \bigskip \bigskip\bigskip
\bigskip\bigskip\bigskip
\bigskip
\bigskip\bigskip\bigskip
\medskip
\medskip
\medskip
\section{Introduction}

In the past few years it has been  demonstrated
that some very simple non-Hermitian Hamiltonians provide well-defined
 quantum theories \cite{Bender:1998kf, Dorey:2001uw}.
These theories are related  via (non-unitary) similarity transformations
\cite{Mostafazadeh:2005wm}
to
`standard' Hermitian quantum theories.
 In most cases it is difficult to determine
  the form of the equivalent Hermitian theory.
In this paper a  non-Hermitian form of quantum electrodynamics (QED)
is considered. 
A non-hermitian, but ${\cal PT}$-symmetric, form of QED
was proposed by Bender and Milton \cite{Bender:1999et}.
Later, Milton \cite{Milton:2003ax}
argued that, due to an axial anomaly, the theory
is not renormalisable.
However,
he suggested an alternative non-Hermitian  theory using an axial vector field instead of an axial
current.  Milton's theory was further developed
in reference  \cite{Bender:2005zz}. 
The present author \cite{Ford:2008vb}
has argued that this theory 
is a magnetic form of QED in that it describes the interactions
  of spin ${1\over 2}$ particles carrying magnetic rather
than electric charges. 
It is not clear what is the Hermitian field theory (or theories)
that is equivalent to magnetic QED (MQED). In this paper we examine
the possibility that MQED is actually equivalent to QED
so that the similarity transformation between the Hermitian and non-Hermitian
descriptions
 is a form of electric-magnetic duality. This proposal requires a
vanishing bare fermion mass in both theories. This is because
Dirac masses are incompatible with the conservation of magnetic charge much as
Majorana masses spoil the conservation of electric charge.
Both versions of QED have the same set of discrete symmetries
${\cal P}$, ${\cal C}$ and ${\cal T}$ and their
anomalous axial
 currents can be identified.
The conserved 
 currents associated with the global gauge symmetries
of the two theories are, however, not identified.
The electric current of QED can be expressed locally in terms of the Dirac
fermion
fields whereas the magnetic current of MQED is local in the fermion
fields of the non-Hermitian theory. It is argued that both
conserved currents are present in both theories
but the magnetic current is non-local in standard QED and
the electric current is non-local in MQED
(though the non-conserved axial current is local in both descriptions).
Consequently, the particles described by massless 
QED (and MQED) carry both electric and magnetic charge.
The four particles described by a Dirac fermion correspond
to the four possible electric and magnetic 
charge assignments of elementary dyons.

Assuming that QED and MQED are equivalent
and that the physical particles correspond to the field content
yields a scale invariant spectrum. The idea of a scale invariant
version of spinor QED has been developed by
Johnson, Baker and Willey (JBW) \cite{Johnson:1963zz,Johnson:1973pd}.
On the basis of a detailed examination of the ultraviolet 
and infrared properties
of Feynman diagrams entering the photon propagator JBW
 argued
that for special values of the fine structure constant, $\alpha$, 
massless QED could be finite and hence scale invariant.
Here  finite theories
correspond to solutions of an `eigenvalue' condition involving
the bare fine structure constant.
 Unfortunately, it 
 is still not known whether this equation has  non-trivial solutions.
 Even if it does, a  physical interpretation of any finite
 theory is lacking. 
However, Adler \cite{Adler:1972in}
 has observed
that a solution of the eigenvalue equation would
provide a finite QED independent of the number of fermion species.
Consequently,  if
the eigenvalue equation has a single non-trivial solution
then finite spinor QED incorporates the quantisation of electric
charge. 
It is argued that this is  a reflection
of the dyonic nature of the physical states; the JBW eigenvalue
equation is interpreted as a version of
Dirac's charge quantisation condition.

 The outline of this paper is as follows.  
Non-Hermitian QED is reviewed in the next section.
The idea of electric-magnetic duality in QED is
outlined in section 3.
In section 4 it is explained how the JBW constraint on the fine structure
constant can be interpreted as a Dirac-like quantisation condition on the
 electric and magnetic charges of the dyons.
Section 5 includes concluding remarks and some speculation
concerning spin-0 dyons.

\section{Magnetic QED}

Massless QED is based on the Lagrangian
(metric and Dirac matrix conventions are as in Bjorken and Drell \cite{Bjorken:1965})
\begin{equation}\label{qed}
{\cal L}=-{1\over 4}F_{\mu\nu}
F^{\mu\nu}+i\bar\psi\gamma^\mu\partial_{\mu}\psi +e\bar \psi
\gamma^\mu A_{\mu}\psi,
\end{equation}
where $F_{\mu \nu}=\partial_\mu A_\nu-\partial_{\nu}A_{\mu}$.
Here $A_{\mu}$ is a $U(1)$ gauge potential
 and $\psi$ is a Dirac spinor.
The corresponding quantum theory
has a Hermitian Hamiltonian and
is symmetric under parity ${\cal P}$ and time-reversal ${\cal T}$.
 Milton considered the Lagrangian \cite{Milton:2003ax} \footnote{
In \cite{Milton:2003ax} a real representation for 
Dirac spinors was adopted. In this paper, as also in 
 \cite{Bender:2005zz}, a conventional complex representation is assumed.} 
\begin{equation}\label{Milton}
{\cal L}=-{1\over 4}G_{\mu\nu}
G^{\mu\nu}+i\bar\psi\gamma^\mu\partial_{\mu}\psi +ig\bar \psi
\gamma^\mu B_{\mu}\psi,
\end{equation}
with $G_{\mu\nu}=\partial_{\mu}B_{\nu}-\partial_{\nu}B_{\mu}$,
$B_{\mu}$ being an abelian gauge potential, $\psi$ a Dirac spinor
and $g$  a real coupling constant.
 The theory couples a gauge
potential $B_{\mu}$ (assumed to be Hermitian) to the anti-Hermitian
current
$
j_\mu=ig\bar \psi \gamma_\mu \psi$
which renders the Hamiltonian non-Hermitian.
The gauge potential transforms in the usual way
under time reversal, i.e.
\begin{equation}\label{TonB}
{\cal T}B_0({\bf r},t){\cal T}^{-1}=B_0({\bf r},-t),~~~~~~~~~~
{\cal T}{\bf B}({\bf r},t){\cal T}^{-1}=-{\bf B}({\bf r},-t),
\end{equation}
or in a more compact form
\begin{equation}\label{TonB2}
{\cal T}B_\mu({\bf r},t){\cal T}^{-1}=B^\mu({\bf r},-t).
\end{equation}
Under parity
a
 non-standard (pseudovector)  transformation is assumed
\begin{equation}\label{PonB}
{\cal P}B_\mu({\bf r},t){\cal P}^{-1}=-B^\mu(-{\bf r},t).
\end{equation}
The resulting theory is non-Hermitian and ${\cal P}$ and ${\cal T}$
are not symmetries. However, the combined operation
${\cal PT}$ is a symmetry and on this basis
the theory is expected to have a real spectrum.

 An alternative Lagrangian for non-Hermitian QED
was given in \cite{Ford:2008vb}.
This theory has the same set of discrete symmetries as standard QED
and thus appears to possess more discrete symmetry than Milton's theory.
This is misleading since the two non-Hermitian theories are
related by a canonical transformation. 
The mismatch reflects the use of different parity operators;
the two parity operators have the same effect on pure photon
states but on  fermionic states the Milton ${\cal P}$
is equivalent to the ${\cal CP}$ operator of \cite{Ford:2008vb}.
The two theories also `disagree' with respect to time-reversal.
However, it is a feature of this anti-unitary operator that
a ${\cal T}$-symmetric theory can be canonically equivalent to
a non  ${\cal T}$-symmetric theory (a simple example
being the Hamiltonian $H(q,p)=q$ which is canonically equivalent
to $K(Q,P)=P$; the former is ${\cal T}$-symmetric but the latter is
not). 
The alternative Lagrangian reads
\begin{equation}\label{newL}
{\cal L}=-{1\over 4}H_{\mu\nu} H^{\mu\nu}
+i\bar \lambda \gamma^\mu\partial_{\mu}
\lambda +ig \bar\lambda \gamma^\mu V_\mu
 \lambda.\end{equation}
Here $H_{\mu\nu}=\partial_{\mu}V_{\nu}-\partial_{\nu}V_{\mu}$ where
the gauge potential, $V_{\mu}$,
has  unconventional transformations
under both ${\cal T}$ and ${\cal P}$, that is
\begin{equation}\label{TonV}
{\cal T}V_\mu({\bf r},t){\cal T}^{-1}=-V^\mu({\bf r},-t),
~~~~~~~~~
{\cal P}V_\mu({\bf r},t){\cal P}^{-1}=-V^\mu(-{\bf r},t). 
\end{equation}
The spinor field $\lambda$
transforms like a Dirac spinor under proper Lorentz transformations
and time-reversal.
Under parity it transforms as
\begin{equation}\label{Ponlambda}
{\cal P}\lambda_\alpha({\bf r},t){\cal P}^{-1}
 ={P}_{\alpha \beta} \lambda_{\beta}^\dagger(-{\bf r},t),
\end{equation}
where $P_{\alpha \beta}$ denotes the matrix elements of the  Dirac matrix
$i\gamma^0 \gamma^2$
(here it is assumed that
 $\gamma_0=\gamma_0^T$ and $\gamma_2=\gamma_2^T$). This
 is actually the standard form of
the ${\cal CP}$ transformation for Dirac spinor fields.
The theory couples the Hermitian gauge potential,
$V_\mu$,
to the 
current
\begin{equation}
\label{newcurrent}
k_\mu=ig\bar\lambda \gamma_\mu\lambda.\end{equation}
Under ${\cal T}$ and ${\cal P}$
\begin{equation}\label{TPonk}
{\cal T}^{-1}k_\mu({\bf r},t){\cal T}=
-k^\mu({\bf r},-t),~~~~~~
{\cal P}^{-1}k_\mu({\bf r},t){\cal P}=-k^\mu(-{\bf r},t).
\end{equation}
This non-Hermitian theory is symmetric under ${\cal T}$ and ${\cal P}$;
the non-standard transformation properties of $V_\mu$ compensate for those
of $k_\mu$. The field strength,
 $H_{\mu\nu}$, transforms like the Maxwell dual field strength
and satisfies the ${\cal P}$ and ${\cal T}$ symmetric equation of motion
\begin{equation}
\label{eqofmotion}
\partial_{\mu} H^{\mu\nu}=k^\nu.\end{equation}

A gauge-invariant mass term 
\begin{equation}
\label{mass}
{\cal L}_{mass}= -m\bar\lambda \lambda,
\end{equation}
may be added 
to the Lagrangian (\ref{newL}).
This looks like a Dirac mass term but \it physically
\rm
 it is
a Majorana mass. Viewed as perturbations, Dirac and Majorana
masses allow a massless fermion to transform into its ${\cal P}$
and $\cal{CP}$ conjugate, respectively. Due to the switched 
 transformation properties of 
the $\lambda$-spinor under ${\cal P}$ and $\cal{CP}$
a  spin-${1\over2}$ monopole
can have a Majorana mass but not a Dirac mass.
This complements the well known result that a spin-${1\over 2}$
particle carrying electric charge 
may have a Dirac mass but not a Majorana mass.

Returning to the massless theory, the Lagrangian (\ref{newL})
possesses the global
 symmetry
\begin{equation}\label{axial}
\lambda\rightarrow e^{s\gamma_5}\lambda,~~~~~~
\bar\lambda\rightarrow \bar \lambda e^{s\gamma_5},
\end{equation} 
where $s$ is a constant.
This symmetry implies that the current
\begin{equation}\label{axialcurrent}
k_\mu^5=\bar\lambda \gamma_\mu\gamma_5\lambda,
\end{equation}
is conserved at the classical level.
Quantum mechanically  this current  exhibits an
anomalous divergence\footnote{
This is a formal continuation of the QED anomaly equation
with $F_{\mu\nu}$ replaced with $H_{\mu\nu}$
and $e$ replaced with $ig$.
}
\begin{equation}\label{nonhermanomaly}
\partial^\mu k_\mu^5=-{g^2\over{8\pi^2}}
H^{\mu\nu}\tilde H_{\mu\nu},\end{equation}
where $\tilde H^{\mu\nu}$ is the Hodge dual of $H^{\mu\nu}$.

Although MQED has a non-Hermitian Hamiltonian
it is expected that it can be brought into to a Hermitian 
form via a (non-unitary) similarity transformation
\begin{equation}\label{similarity}
S^{-1}H_{MQED}S=h ~~~~~\mbox{ with } ~~~~~
h^\dagger=h.
\end{equation}
The Dirac inner-product $(\psi,\phi)_D=\langle \psi|\phi\rangle$
is not invariant under such similarity transformations.
The choice of the Dirac inner-product within the hermitian theory
is equivalent to a modified inner-product for the non-Hermitian theory
\begin{equation}\label{newinner}
(\psi,\phi)=\langle \psi|\eta|\phi\rangle,
\end{equation}
where $\eta=(S^\dagger S)^{-1} $.
Instead of focussing on the similarity transformation, $S$, 
one may take  (\ref{newinner}) as a starting point, with a view to determining a suitable
$\eta$ operator.
In particular, the Dirac inner product does not provide unitary time-evolution
for a non-Hermitian Hamiltonian.
 The inner-product (\ref{newinner}) does provide this
if
\begin{equation}\label{unitary}
\eta H-H^\dagger \eta=0.\end{equation}
In general, if $U$ belongs to the symmetry group of the theory
(including discrete as well as continuous symmetries)
one has
\begin{equation}\label{symmetry}
 U^\dagger\eta U=\eta.\end{equation} 
Equation (\ref{unitary}) is (\ref{symmetry}) applied to time translations.
 In MQED, as well as other non-Hermitian theories,
 it is straightforward to identify an operator $\eta$ fulfilling
equation (\ref{unitary}). Unfortunately, a solution to
 equation (\ref{unitary}) on its own is not sufficient to
guarantee  that the associated inner-product is the physical one.
In reference \cite{Bender:2005zz} a perturbative expansion for
 $\eta$ 
is developed.

\vfill\eject

\section{Electric-Magnetic Duality in QED}

Assuming that the fields in QED correspond to the physical
particles, the spectrum is based on photons and four fermions
derived from the Dirac spinor field $\psi$.  In massless QED the one 
fermion states
comprise  left and right handed electrons and their antiparticles:
\medskip

A) left-handed electron

B) right-handed anti-electron

C) right-handed electron

D) left-handed anti-electron

\medskip\noindent
These particles can be
 transformed into each other 
through the discrete symmetries ${\cal P}$, ${\cal CP}$
and ${\cal C}$.

\begin{figure}
\setlength{\unitlength}{1mm}
\begin{picture}(100,80)

\put(45,13){\vector(1,0){24}}
\put(45,13){\vector(-1,0){24}}
\put(45,61){\vector(1,0){24}}
\put(45,61){\vector(-1,0){24}}

\put(13,37){\vector(0,1){18}}
\put(13,37){\vector(0,-1){18}}

\put(77,37){\vector(0,1){18}}
\put(77,37){\vector(0,-1){18}}

\put(45,64){\makebox(0,0){${\cal P}$}}
  \put(45,10){\makebox(0,0){${\cal P}$}} 

\put(81,37){\makebox(0,0){${\cal CP}$}}
  \put(9,37){\makebox(0,0){${\cal CP}$}}

 \put(14,14){\makebox(0,0){$B$}}
 \put(14,60){\makebox(0,0){$A$}}
 \put(76,60){\makebox(0,0){$C$}}
 \put(76,14){\makebox(0,0){$D$}}

\end{picture}
\caption{The four particles are related by the discrete symmetries ${\cal P}$ and ${\cal CP}$}
\end{figure}
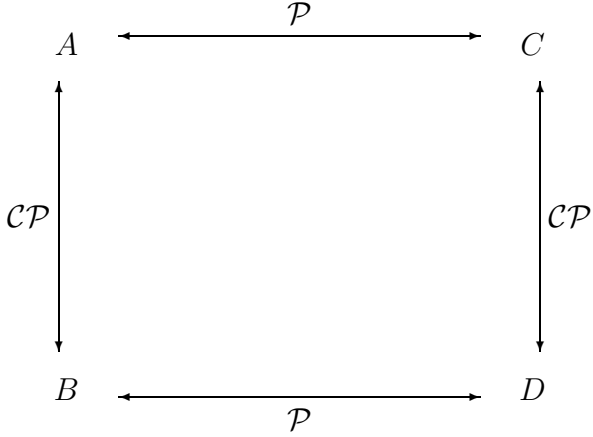

 Now assume the existence of (non-negative)
  commuting
 number operators $N^A$, $N^B$, $N^C$ and $N^D$
for the  particle types listed above.
Under ${\cal P}$ and  ${\cal CP}$ 
they transform according to Fig.1, for example, ${\cal P}^{-1}N^A{\cal P}=N^C$.
The number operators are invariant under time-reversal, so that
${\cal T}^{-1}N^A{\cal T}=N^A$, etc.
In QED a construction of such operators
 is lacking (except for the free theory).
However, operators are known which can be interpreted as two
particular linear
 combinations:
 \begin{equation}\label{electriccharge}
Q_e =e(N^A-N^B+N^C-N^D),\end{equation}
and
\begin{equation}\label{axialcharge}
Q_5=N^A-N^B-N^C+N^D.
\end{equation}
The electric charge, $Q_e$, and axial charge, $Q_5$, are defined
as volume  integrals of the 
local currents $j_\mu=e\bar \psi \gamma_\mu\psi$.
and  $j^5_\mu=\bar \psi \gamma_\mu\gamma_5\psi$,  respectively.
Electric charge is conserved whereas the axial current
 satisfies
the anomalous divergence condition \cite{Adler:1969gk,Bell:1969ts,Adler:1969er}
\begin{equation}\label{anomaly}
\partial^\mu j_\mu^5={\alpha\over{2\pi}}F^{\mu\nu}\tilde F_{\mu\nu},
\end{equation}
where $\alpha=e^2/4\pi$ is the bare fine structure constant.

Comparing
 massless QED with massless MQED one has two quantum theories with the same
set of discrete symmetries and the same number of spinor and vector
degrees of freedom.
Can the
 two theories be equivalent?
This would mean that the two Hamiltonians would be related by a
similarity transformation of the form (\ref{similarity})
with $h=H_{QED}$.
  If this holds  how do the operators or degrees of freedom
match?
At the
 operator level it is natural to identify the anomalous currents, 
\begin{equation}\label{dualcurrents}
S^{-1}k_\mu^5 S=j_\mu^5
\end{equation} 
 since
they have
 the same transformation properties with respect to the discrete symmetries.
Moreover, the associated axial
 charges are expected to have integer
eigenvalues. The anomaly equations then give
\begin{equation}\label{anomalymatch}
-g^2 S^{-1}H^{\mu\nu}\tilde
H_{\mu\nu}S=e^2F^{\mu\nu}\tilde F_{\mu\nu}.\end{equation}
which is consistent with the identifications
\begin{equation}\label{HisFdual}
S^{-1}H_{\mu\nu}S=^{\!\!\!\! ?}\tilde F_{\mu\nu},
\end{equation}
and
\begin{equation}g=\pm e.\label{selfdual}
\end{equation}
The identification (\ref{HisFdual}) is too strong to hold for all matrix elements;
it is only expected to hold with respect to photon states.\footnote{
Assuming a finite QED and MQED
we have $T^{00}={1\over 2}(E^2+B^2)+
\hbox{ fermion bilinear }$ in both theories.
Pure photon states are annihilated by the
fermion bilinear.
Therefore, on photon states the constant of proportionality
between $S^{-1}H_{\mu\nu}S$ and $\tilde F_{\mu\nu}$
must be $\pm 1$ which is the basis for
(\ref{selfdual}).
A loophole
in the argument is that
$F^{\mu\nu}\tilde F_{\mu\nu}$ annihilates pure photon
 states. }
 One can understand this at the classical level. 
In the presence of magnetic charges $F_{\mu\nu}$
requires Dirac strings whereas for $H_{\mu\nu}$
string singularities are attached to electric charges.
Therefore $\tilde F_{\mu\nu}$ and $H_{\mu\nu}$ 
cannot be identical in the presence of charges.

The conserved currents associated with the global gauge
symmetries cannot be identified as they transform differently
under ${\cal P}$, ${\cal CP}$ and ${\cal T}$. For example, under parity
\begin{equation}\label{Poncurrents}
{\cal P}^{-1}j_\mu({\bf r},t){\cal P}=
j^\mu(-{\bf r},t),~~~~~~~~~
{\cal P}^{-1}k_\mu({\bf r},t){\cal P}=-k^\mu(-{\bf r},t).
\end{equation}
Now consider the charge
\begin{equation}\label{magneticcharge}
Q_m=ig(N^A+N^B-N^C-N^D),~~~~g=\pm e.\end{equation}
   This operator cannot be expressed locally in the QED variables.
   However, it has the same transformation properties as the
   magnetic charge operator
    in MQED (defined as the volume integral of $k_0$).
Under the proposed duality massless
QED possesses an \it additional \rm
conserved current
\begin{equation}
j_{\mu}^{m}(x)=S^{-1}k_\mu(x)S.\end{equation}
It is natural to identify
$Q_m$ as a volume integral of $j_{m}^0$.
Consequently, massless QED
has an additional  symmetry which has no classical analogue.
One can say that this symmetry `replaces' chiral symmetry which,
by virtue of the axial anomaly,
does not survive quantisation.
Although the symmetry groups of classical and quantum electrodynamics
are different they have the same dimension.

We have argued that scale invariant QED and MQED
describe the same physics.
   Accordingly, the one particle states carry electric
 \it and \rm magnetic charge;
   a state $|A\rangle$ which is an eigenvector of $N^A$ with unit eigenvalue
   has the properties
   \begin{equation}\label{stateAcharges}
   Q_e|A\rangle=e|A\rangle,~~~~~~~
   Q_m|A\rangle=ig|A\rangle,
\end{equation}
and similarly for one particle states $|B\rangle$, $|C\rangle$ and 
$|D\rangle$ which are ${\cal CP}$-, ${\cal P}$- and  ${\cal C}$- conjugates
of $|A\rangle$, respectively.  
A simple process which exhibits the conservation of electric and magnetic
 charge is
\begin{equation}
A+D\rightarrow B+C.
\end{equation}
Here $\Delta Q_5=-4$.
One can also have annihilation of charge conjugate pairs, e.g.
\begin{equation}
A+D\rightarrow \hbox{photons},
\end{equation}
for which $\Delta Q_5=-2$.

\section{Finite QED and the Quantisation of Electric and
Magnetic Charge}

An electric-magnetic duality between
 QED and  MQED
suggests a spectrum based on photons and massless spin-$\frac{1}{2}$
dyons carrying equal or opposite electric and magnetic charges.
The four particle states associated with a Dirac fermion correspond to the
four possible charge assignments of elementary dyons.
 Therefore if this duality is realised
the relevant quantum field theory is scale-invariant. 
The idea that a finite scale-invariant version of spinor QED
may exist
was advanced by Johnson, Baker and Willey (JBW) in the early
1960s.  
Starting from an integral representation of the photon propagator 
JBW  argued that QED is finite\footnote{
The possibility of a finite photon propagator was considered
earlier by Gell-Mann and Low \cite{GellMann:1954fq}. JBW
went further by positing the finiteness of full QED.}
 if
$\alpha$ satisfies the `eigenvalue' condition
\begin{equation}\label{JBW}
f(\alpha)=0,
\end{equation} 
where $f(\alpha)$
is a coefficient in the large-momentum expansion of the integrand.
Unfortunately, it is still not known
for what positive
values of $\alpha$, if any\footnote{It has been suggested
that (\ref{JBW}) has no positive solution,
see for example \cite{Acharya:1996qz}.},
equation
 (\ref{JBW}) holds. Although $f(\alpha)$ has been analyzed
 perturbatively
the determination of the roots of $f$ is a non-perturbative
problem.
Adler \cite{Adler:1972in}
 has observed that a solution of (\ref{JBW}) would provide a finite QED
independent of the number of fermion species.
In other words if (\ref{JBW}) has a single positive solution
then finite spinor QED incorporates the quantisation
of electric charge.
We would like to interpret this quantisation condition,
and hence the JBW eigenvalue equation,
as a Dirac-like quantisation condition
on the electric and magnetic charges of the dyons.
Dirac's original quantisation condition \cite{Dirac:1931}, constraining the charge of
an electron interacting with a  monopole, does not apply
to dyons. Schwinger and Zwanziger (SZ) have argued that
the possible electric and magnetic charges
of any charged particles are constrained by the quantisation
condition \cite{Schwinger:1968rq,Zwanziger:1968rs}
\begin{equation}\label{SZ}
e_ig_j-g_ie_j=2\pi n_{ij},\end{equation}
 where $e_i$ and $g_i$ denotes the electric and magnetic charge
 of the $i$th particle, respectively.
 $n_{ij}$ is an even integer\footnote{
Dirac's quantisation condition allows the $n_{ij}$ to be odd.
This corresponds to quantising the angular momentum of the electromagnetic
fields in units of $\frac{1}{2}\hbar$.  As photons are spin-$1$ particles
a restriction to even $n_{ij}$ is natural.}.
Applying this quantisation condition to
the proposed dyon spectrum  of QED yields
\begin{equation}\label{value}
\alpha={1\over 2},
\end{equation}
assuming that the $n_{ij}$
are restricted to the minimal values
$2$, $-2$ and $0$. 
 While this may turn out to be the `correct' value of the fine structure
constant,
the use of the SZ quantisation condition is questionable
 in this instance (see also     \cite{Ross:1987ia}).
Basic assumptions 
underlying the SZ derivation of the dyon quantisation condition
are contrary to those of this paper. These differences can be
summarised as follows:

\medskip

i) As in this paper SZ consider Lagrangians/Hamiltonians for
spin-$\frac{1}{2}$ particles carrying magnetic charge.
Both approaches involve a gauge potential, $V_\mu$,
with the transformation properties (\ref{TonV}).
However, the SZ
 Hamiltonians are Hermitian and also
break ${\cal P}$ and ${\cal T}$.
\medskip

ii) To describe dyons SZ employ Lagrangians with two 
gauge potentials, $A_\mu$ and $V_\mu$.
As the associated field strengths
 are dual,  $V_\mu(x)$ can be considered as a functional
of $A_\nu$ or vice versa. Therefore these Lagrangians
 are non-local. In this paper dyons are interpreted as the
 states arising from the
quantisation of the local QED Lagrangian or the local MQED Lagrangian.
\medskip

The Hamiltonians considered by SZ are tied to 
  a generalisation of the Lorentz force law
proposed by Dirac \cite{Dirac:1948}
wherein a particle carrying an electric charge $e$ and magnetic charge
$g$ is governed by the equation of motion
\begin{equation}
\label{Diracforce}
m{d^2x_\mu\over{d\tau^2}}=\left(eF_{\mu\nu}+g\tilde F_{\mu\nu}\right)
{dx^\nu\over{d\tau}},\end{equation}
where $m$ is the mass of the particle and $\tau$ denotes proper-time. The  field strength tensor, $F_{\mu\nu}$,
and its Hodge dual, $\tilde F_{\mu\nu}$, obey the generalised Maxwell equations
\begin{equation}\label{Maxwell}
\partial_\mu F^{\mu\nu}=j^\nu,~~~~~~~~~~
\partial_\mu\tilde F^{\mu\nu}=j_{mag}^\nu.
\end{equation}
Electric and magnetic charges are defined
as integrals (over a volume containing the charge)
of $j^0$ and $j^0_{mag}$ , respectively.
 Non-hermitian forms of QED cannot, at least for weak coupling,
 produce a force law of the form (\ref{Diracforce}); the  coupling of  an anti-hermitian
 current to the gauge potential gives rise to an \it attractive
\rm  force between like charges \cite{Milton:2003ax}. 

Maxwell's equations (\ref{Maxwell}), the Dirac force law
and the SZ quantisation condition
  are invariant under the Heaviside
duality rotation
\begin{equation}\label{Heaviside}
{\bf E}\rightarrow \cos\theta ~{\bf E}+\sin\theta ~{\bf B},
~~~~~{\bf B}\rightarrow\cos\theta ~{\bf B}-\sin\theta~ {\bf E},
\end{equation}
\begin{equation}\label{Heavisidecurrent}
j^\mu\rightarrow \cos \theta {j^\mu}+\sin \theta j^\mu_{mag},~~~~~
j^\mu_{mag}\rightarrow \cos\theta  ~j^\mu_{mag}-\sin\theta~
j^\mu,
\end{equation}
\begin{equation}\label{Heavisidecharges}
e_i\rightarrow \cos\theta~ e_i+\sin\theta~ g_i,~~~~~
g_i\rightarrow \cos\theta~ g_i-\sin\theta ~e_i,
\end{equation}
where $\theta$ is a constant.
It is clear that there is no room for the transformation
(\ref{Heavisidecharges}) in finite QED; the electric charges are fixed by the 
eigenvalue condition.
As our approach is based
on the idea of electric-magnetic duality it may seem
contradictory to reject Heaviside duality.
The electric-magnetic duality
we are considering is more akin to Olive-Montonen (OM)
duality \cite{Montonen:1977sn}.   Here the interactions of dyons, photons and other particles
 are described by two distinct quantum field theories
 - one theory admits a perturbative expansion in the electric charge(s)
 the other a
 perturbative
 expansion in
 the magnetic charge(s) but non-perturbatively they are equivalent. 
OM duality relations have been conjectured
for  certain supersymmetric Yang-Mills-Higgs theories.
These theories are  Hermitian and are also expected to be finite.

If the SZ quantisation condition does not apply here, what should replace it?
A straightforward proposal is to determine the roots of $f$.
If there is a single positive root then  this statement on its own
would suffice to replace (\ref{SZ}).
Unfortunately, there is not a clear path to tackle equation (\ref{JBW}).
In the case of the Dirac quantisation condition
there are a number of intuitive derivations \cite{Wilson:1949zz, Goldhaber:1965} which reproduce
Dirac's result; using a mixture of classical and quantum ideas
 quantisation conditions can
 be deduced without formulating the full quantum theory.
  These derivations assume massive, or even infinite mass,
 particles whereas we seek a quantisation condition for massless
 dyons. Nevertheless, given that  a solution of (\ref{JBW}) is not within reach,
 an informal
 derivation may provide a hint towards a complete solution.
 An approach to the Dirac quantisation condition
 is to consider the motion of an electron 
 in the presence of a static monopole \cite{Goldhaber:1965};
 one finds that the change of a component of the orbital angular momentum,
during deflection by the monopole,
is always $eg/2\pi$.
Assuming that this is quantised in units of $\hbar$
 gives Dirac's result.
 To repeat this exercise for dyons one
 requires a generalisation of the Lorentz
 force law to accommodate magnetic charges.
 Using the Dirac force law      leads to a SZ-type quantisation rule.
   However,
 the Dirac force law
 does not fit
 with MQED since it  together with (\ref{Maxwell})
 gives a repulsive force between like magnetic charges.
 A generalistion of the Lorentz force law
 that incorporates an attractive force between like          charges           is \cite{Ford:2008vb}
 \begin{equation}\label{newforce}
m{d^2x_\mu\over{d\tau^2}}=\left(e F_{\mu\nu}-g\tilde F_{\mu\nu}\right)
{dx^\nu\over{d\tau}},\end{equation}
where the generalised Maxwell equations (\ref{Maxwell}) are unchanged.
A study of the classical dynamics of dyons under this      force law
might be instructive.

\section{Concluding Remarks}

In this paper we have considered the possibility of an Olive-Montonen
type duality between massless QED and a non-hermitian version of massless QED.
The proposal is based on the equivalence of the discrete symmetries
and the observation that the properties of the anomalous axial currents match.
As the quantum theory is simultaneously described
by an electric and magnetic version of QED
physical particles
carry both electric and magnetic charge.
That is, massless QED is a theory of interacting dyons.
Accordingly,
a Dirac-like quantisation condition is expected.
We have argued that this
 corresponds to the JBW eigenvalue condition
 developed in studies of finite spinor QED.
 This interpretation
 is consistent with
 Adler's observation concerning
 species independence.

The interpretation of massless QED as a theory of dyons appears not to extend
to  scalar electrodynamics - one complex scalar field cannot be identified
with four elementary dyons.
This fits with suggestions that the JBW program is not tenable for scalar QED \cite{FlammandFreund1964,Fry:1973qf}.
Does this mean that there is no relativistic quantum theory of abelian spin-0 dyons?
There is, however, a conceptually simple  though speculative,
way to obtain spinless dyons from \it spinor \rm
QED. Hitherto we have assumed that
one particle states correspond
to the field content of the theory.
 In a strongly
coupled theory this is possible but by no means mandatory, e.g.
the spectrum of pure       Yang-Mills theory
does not match its gluon fields.
In section 3 it was assumed that one-particle states
are eigenvectors of the number operators with one unit and three zero
 eigenvalues.
Instead consider states 
with eigenvalue $\frac{1}{2}$;
a state $|a\rangle$ is assumed to have the properties
\begin{equation}\label{stateanumbers}
N^A|a\rangle=\frac{1}{2}|a\rangle
,~~~~~~N^B|a\rangle=N^C|a \rangle=N^D|a\rangle=0.\end{equation}
Similarly one can define states $|b\rangle$, $|c\rangle$ and $|d\rangle$
which, much like the one fermion states, are related by the discrete symmetry
operations, 
e.g. $|b\rangle={\cal CP} |a\rangle$.  These states carry fractional
electric \it and \rm  magnetic charge
\begin{equation}\label{stateacharges}
Q_e|a\rangle=\frac{1}{2}e|a\rangle,~~~~~~
Q_m|a\rangle=\frac{i}{2}g|a\rangle.\end{equation}
The idea that fermionic theories possessing a charge-conjugation symmetry
allow fractionally charged states goes back to the work of 
Jackiw and Rebbi\footnote{These authors 
performed a semi-classical analysis
of fractionally charged states
via fermion zero modes.
It is not clear whether the existence
of such modes is strictly necessary.
In fact, fermion zero modes 
do exist
for certain abelian dyon backgrounds
on $R^3$ \cite{VanBaal:2002rt}.} \cite{Jackiw:1975fn}
who argued that
such states would be spinless.
If this scenario is realised for finite $\alpha$ 
it is possible that $\alpha$ is not a solution of the eigenvalue 
condition which derives from the finiteness of 
the photon propagator. If the charged states are not associated with the fermion
fields it is possible that photons are not physical states.
Moreover, as the fractionally charged states carry
half the electric and magnetic charges of the fields 
a different value of $\alpha$ may be required
to allow a consistent quantisation.
\vfill\eject

\end{document}